\documentclass{article}

\usepackage{arxiv}

\usepackage{cite}
\usepackage{amsmath,amssymb,amsfonts,mathrsfs}
\usepackage[ruled,linesnumbered]{algorithm2e}
\usepackage{algorithmic}
\usepackage{graphicx}
\usepackage{textcomp}
\usepackage{xcolor}
\usepackage{subcaption} 
\usepackage{cleveref,verbatim}

\title{Cost-Efficient Computation Offloading and Service Chain Caching in LEO Satellite Networks}
\date{}

\author{Yantong~Wang,~Chuanfen Feng, Jiande Sun\\
	School of Information Science and Engineering\\
	Shandong Normal University\\
	Ji'nan, 250358, China\\
	\texttt{\{yantong,fengcf,jiandesun\}@sdnu.edu.cn} \\
}

\renewcommand{\shorttitle}{\textit{arXiv} Template}

\begin{document}
\maketitle

\begin{abstract}
	The ever-increasing demand for ubiquitous, continuous, and high-quality services poses a great challenge to the traditional terrestrial network. To mitigate this problem, the mobile-edge-computing-enhanced low earth orbit (LEO) satellite network, which provides both communication connectivity and on-board processing services, has emerged as an effective method. The main issue in LEO satellites includes finding the optimal locations to host network functions (NFs) and then making offloading decisions. In this article, we jointly consider the problem of service chain caching and computation offloading to minimize the overall cost, which consists of task latency and energy consumption. In particular, the collaboration among satellites, the network resource limitations, and the specific operation order of NFs in service chains are taken into account. Then, the problem is formulated and linearized as an integer linear programming model. Moreover, to accelerate the solution, we provide a greedy algorithm with cubic time complexity. Numerical investigations demonstrate the effectiveness of the proposed scheme, which can reduce the overall cost by around 20\% compared to the nominal case where NFs are served in data centers.
\end{abstract}


\section{Introduction}

Recently, network access has been available to the majority of the world population \cite{ericsson2023report}. However, more than half of the earth's space, especially for remote or unreachable areas like rural regions, deserts, volcanoes, and oceans, still lacks network coverage \cite{yu2022ecsagins}. Due to the massive costs behind infrastructure deployment in the aforementioned places and the vulnerability to natural disasters, traditional terrestrial networks cannot meet the increasing demand for ubiquitous and continuous network services. As an indispensable supplement to the ground network, low earth orbit (LEO) satellites break the geographic limitations and provide wide space coverage, which attracts attentions from both industry and academia\cite{mohammad2022evolution,zhou2023aerospace}. 
Furthermore, to extend the role of satellites and bring computation resources closer to the remote areas, the mobile/multi-access edge computing (MEC) technology has been introduced into the satellite system \cite{xie2020satellite}. Therefore, LEO satellites can not only support network connectivity but also provide on-board computing and caching services. 

A task computation requires both input data and program code. Considering applications in remote regions, for instance, the environment monitoring, the time-sensitive input data is hardly reusable for the next-time executions. However, the program for data processing is utilized frequently\cite{bi2020twc_joint}. Moreover, with the assistance of software-defined networking (SDN) and network function virtualization (NFV) technologies, LEO satellite can manage network resources efficiently and flexibly, in which network functions (NFs) are orchestrated in a certain predefined order to establish a service chain. Additionally, owing to the limitations on the energy and hardware resources in terminals, local computation tasks, especially those with heavy workload, should be processed on either LEO satellites hosting the required NFs, or the terrestrial data center with extra transmission latency.
Our aim is to make two tightly coupled decisions in LEO satellite networks, i.e. \textit{where to cache service functions} and \textit{where to offload computing tasks}.

The computation offloading and caching decisions for satellite systems have attracted significant attention in prior art. The authors in \cite{wang2018computation} propose a computation offloading strategy with double edge computing architecture. A three-tier networks with hybrid cloud and edge computing LEO satellite is studied in \cite{song2021energy} for energy-efficient computation offloading. The work in \cite{yang2022cost} introduces a threshold-based method to determine offloading mode for satellite edge computing systems. In \cite{tang2021iotj_computation}, the alternating direction method of multipliers is utilized to minimize energy consumption. The authors in \cite{zhang2023satellite} employs a deep reinforcement learning algorithm to allocate computation tasks. However, the above researches \cite{wang2018computation,song2021energy,yang2022cost,tang2021iotj_computation,zhang2023satellite} make offloading decisions with assumptions that required functions are available in all satellites and services cannot be partitioned. Therefore, these solutions may not scale well in the considered chain caching cases. Regarding the NF placement decisions, an integer non-linear programming model is proposed in \cite{wang2020sfcbased} to minimize resource cost. The study in \cite{yue2023delay} transforms the caching decision problem into weighted graphs. However, the communication and cooperation via intersatellite link (ISL) are not fully considered in the most existing works, and, thus, not make full use of available resources in satellites.

To complement the related works, in this paper we jointly consider the design of collaborative computation offloading and service chain caching in LEO satellites networks, which is formulated and linearized as an integer linear programming (ILP) model. The network resource limitations and the specific task operation order in a service chain are also considered. To accelerate the solution, we also provide a greedy-based method with cubic time complexity. Numerical investigations reveal that the proposed model and greedy algorithm can make better decisions for caching and computation offloading when comparing to other widely-used algorithms.

\section{System Model}
\label{sec:model}

\subsection{Network Architecture}
\label{sec:sec2_part1}
\begin{figure}[htbp]
	\centering
	\includegraphics[width=.8\textwidth]{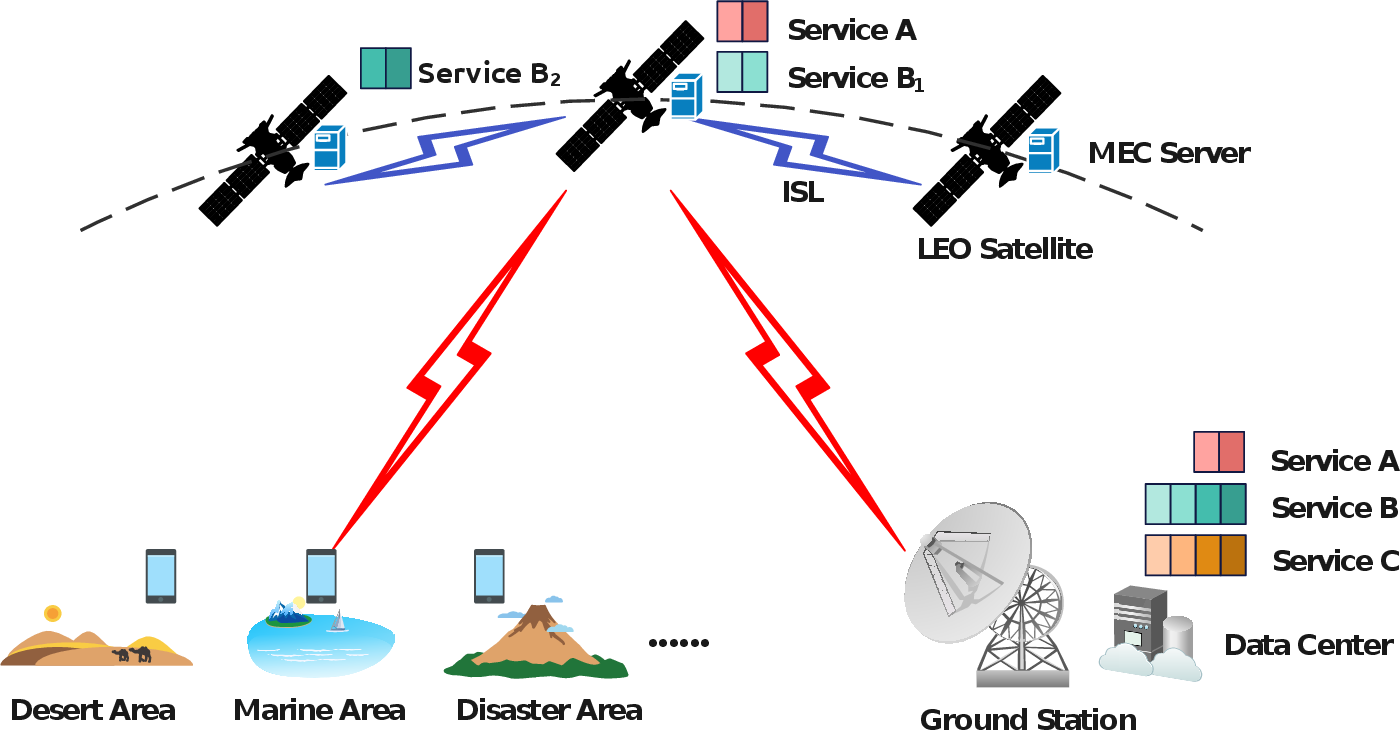}
	\caption{System Architecture}
	\label{fig:structure}
\end{figure}

As illustrated in Fig.\ref{fig:structure}, the network is modelled as an undirected graph $\mathcal{G}=\{\mathbf{V},\mathbf{L}\}$, where $\mathbf{V}$ denotes the set of vertices and $\mathbf{L}$ is the set of links. $\mathbf{V}$ is combined with terminal set $\mathbf{U}$, LEO satellite set $\mathbf{S}$, and ground station set $\mathbf{G}$. In this article, we focus on the scenario that terminals locate in remote or harsh environment, and thus, receive network service via LEO satellite instead of terrestrial communication systems. Due to the local constraint on energy and hardware resources, the computation tasks should be offloaded to the LEO satellite, which is equipped with an edge server and hence provides on-board computation services, or the data center\footnote{We consider the scenario that NFs can be cached in LEO satellite and data center instead of terminals. Nevertheless, local computing mode can be applied with small changes to our model by adding the related delay cost and energy consumption in eq.\eqref{eq:delay_total} and eq.\eqref{eq:energy_total}, respectively.}. The offloading and caching decisions are determined in a controller, which can be established on medium earth orbit (MEO) satellites or geostationary earth orbit (GEO) satellites\cite{qiu2019deep}. Additionally, the available resource of storage space and computation capability on satellite $s$ are represented by $L_s$ and $C_s$ respectively. The data center collocates with ground stations is assumed to have sufficient resource to support service caching and computation offloading\footnote{The term \textit{ground station} and \textit{data center} will be used interchangeably.}. As a result, all of the services are available in the data center. For the sake of simplicity, we assume there are $\vert\mathbf{U}\vert$ terminals, $\vert\mathbf{S}\vert$ satellites, and one ground station in the system.

The entire system operates in a time slot mode. Service caching allocations and computation offloading decisions are determined slot by slot. 
Similar to the existing research\cite{tang2021iotj_computation}, we consider a quasi-static network model, i.e. the network topology, the associations of terminals, and the states of communication channels are fixed during one-slot time window. Hereafter we concentrate on optimization procedures in one time epoch.

For the purpose of network modelling and without loss of generality, we assume each terminal $u$ is associated with a single service request $j\in\mathbf{J}$ in a time slot, which is presented by a binary indicator $\mathbb{R}_{uj}\in\{0,1\}$. Moreover, we use $\mathbb{A}_{us}\in\{0,1\}$ to illustrate the connection between terminal $u$ and satellite $s$.  
Each service $j$ can be partitioned into several NFs $k\in\mathbf{K}$ with a specific order and the NF can be cached on each satellite via integrated edge server. $I$ is utilized to represent the number of NFs in a service chain. The computation resource to execute one bit and caching space requirement of NF $k$ are represented by $c_k$ and $l_k$ respectively. In addition, a binary indicator is utilized with $\mathbb{V}_{ijk}=1$ representing that the $i^{\text{th}}$ function in service $j$ is NF $k$, otherwise, we have $\mathbb{V}_{ijk}=0$. In this paper, we assume that there is no duplicate of NF $k$ in the service chain $j$, i.e. $\sum_{i}\mathbb{V}_{ijk}\leq 1$. Furthermore, $l_{uij}$ is defined as the input data size of the $i^{\text{th}}$ function in service $j$ from terminal $u$. Similar to~\cite{zhang2023satellite}, the output data size of the last NF, i.e. $l_{uIj}$, is relatively small compared with the input data size. As a result, we omit the transmission cost of the return trip.

In this paper, the decisions of NF caching and computation offloading are considered jointly. Therefore, the following decision variable sets are introduced:
\begin{itemize}
	\item \textbf{NF Caching}: $\mathbf{X}\!=\!\{x_{ks}|x_{ks}\in\{0,1\},\! k\in\mathbf{K},\!s\in\mathbf{S}\}$, where $x_{ks}=1$ indicates that NF $k$ is cached in satellite $s$, otherwise, we have $x_{ks}=0$.
	\item \textbf{Computation Offloading}: $\mathbf{Y}\!=\!\{y_{uks}|y_{uks}\in\{0,1\},\! u\in\mathbf{U},\!k\in\mathbf{K},\!s\in\mathbf{S}\}$ with $y_{uks}=1$ representing that the request of NF $k$ from terminal $u$ is served by satellite $s$, otherwise, $y_{uks}=0$.
\end{itemize} 
The key notations used in this paper are summarized in Tab.~\ref{tab:notations}. 

\begin{table}[htb]
	\centering
	\caption{Summary of Main Notations}
	\label{tab:notations}
	\begin{tabular}{cl}
		\hline
		\textbf{Notation}&\textbf{Definition}\\
		\hline
		$\mathbf{U}$&Set of terminals\\
		$\mathbf{S}$&Set of satellites\\
		$\mathbf{G}$&Set of ground stations\\
		$\mathbf{K}$&Set of network functions\\
		$\mathbf{J}$&Set of services\\
		$\vert\mathbf{X}\vert$&The cardinality of set $\mathbf{X}$ \\
		\hline
		$\mathbb{R}_{uj}$&Indicator of service $j$ if it is required by terminal $u$\\
		$\mathbb{A}_{us}$&Indicator of terminal $u$ if it accesses network via satellite $s$\\
		$\mathbb{V}_{ijk}$&Indicator of NF $k$ if it is the $i^{\text{th}}$ function in service $j$\\
		\hline
		$c_k$&The required CPU cycles to execute one bit in NF $k$\\
		$f_{ks}$&The computation capability allocated to NF $k$ in satellite $s$\\
		$f_{g}$&The computation capability allocated to NF in data center\\
		$C_s$&The computation capability of the edge server in satellite $s$\\
		$l_k$&The required storage space of NF $k$\\
		$L_s$& The storage capacity of the edge server in satellite $s$ \\
		$l_{uij}$&The input data size of the $i^{\text{th}}$ NF in service $j$ from terminal $u$ \\ 
		$r_u$,$p_u$&The uplink capacity and transmission power of terminal $u$\\
		$r_s$,$p_s$&The data rate and one-hop transmission power of ISL\\
		$r_g$,$p_g$&The channel capacity and transmission power of satellite-to-terrestrial downlink\\
		$d_{u}$,$d_{g}$,$d_{s}$&The distance between terminal and associated satellite, satellite and data center, and adjacent satellites\\
		$h_{ss'}$&The number of bops between satellite $s$ and $s'$\\
		\hline
		$x_{ks}$&Decision variable indicates whether NF $k$ is cached in satellite $s$\\
		$y_{uks}$&Decision variable indicates whether the request of NF $k$ from terminal $u$ is served by satellite $s$\\
		\hline
	\end{tabular}
\end{table}

\subsection{Latency Cost Model}
The end-to-end latency includes four components: queueing, transmission, propagation, and computing delay, which is analyzed as follows.

\subsubsection{Terminal to Satellite ($\mathbf{U}\rightarrow \mathbf{S}$)}
Based on the notations introduced in~\ref{sec:sec2_part1}, we use $\sum_{s}\sum_{j}\mathbb{A}_{us}\mathbb{R}_{uj}l_{u1j}$ to represent the transmitted data size. Moreover, let $r_{u}$ be the communication capacity of the uplink between terminal and satellite. Then, the transmission latency can be expressed as 
\begin{equation}
	\label{eq:delay_Tus}
	T^{T(\mathbf{U}\rightarrow \mathbf{S})}_u=\frac{\sum_{s\in\mathbf{S}}\sum_{j\in\mathbf{J}}\mathbb{A}_{us}\mathbb{R}_{uj}l_{u1j}}{r_{u}}
\end{equation} 
It is worth noting that unlike the communication in cellular networks, the propagation delay in satellite networks cannot be ignored due to the large distance. As aforementioned, we concentrate on the computation offloading decisions on LEO and date center. Therefore, the local computing delay is not included. Furthermore, with the assumption that each terminal requests a single service in one time slot, there is no queueing delay on the terrestrial-to-satellite link. As a result, the entire roundtrip latency becomes
\begin{equation}
	\label{eq:delay_RTus}
	T^{RT(\mathbf{U}\rightarrow \mathbf{S})}_u=	T^{T(\mathbf{U}\rightarrow \mathbf{S})}_u+\frac{2d_{u}}{c}
\end{equation}
where $d_{us}$ illustrate the distance between terminal $u$ and satellite $s$, and $c$ represents the light speed.

\subsubsection{Satellite to Satellite ($\mathbf{S}\rightarrow\mathbf{S'}$)}Then, we employ $r_s$ as the transmission rate among satellites via ISL. When satellite $s$ does not host the required NF $k$, the computation task should be offloaded to either cached satellite $s'$ or ground station $g$. For the case of offloading to satellite $s'$, let $h_{ss'}$ be the number of hops between satellite $s$ and $s'$. Notably $h_{ss'}=0$ if $s=s'$. Thus, the transmission delay is
\begin{equation}
	\label{eq:delay_TISL}
	\begin{aligned}
 T^{T(\mathbf{S}\rightarrow\mathbf{S'})}_u\!=\!\sum_{j\in\mathbf{J}}\sum_{k,k'\in\mathbf{K}}\sum_{s,s'\in\mathbf{S}}\mathbb{R}_{uj}\Bigl(l_{u1j}\mathbb{A}_{us}\mathbb{V}_{1jk}\cdot y_{uks'}+\sum_{i=1}^{I-1}l_{u(i+1)j}(\mathbb{V}_{ijk}y_{uks})(\mathbb{V}_{(i+1)jk'}y_{uk's'})\Bigr)h_{ss'}\frac{1}{r_s}
	\end{aligned}
\end{equation}
The first term with subscript $i=1$ in eq.\eqref{eq:delay_TISL} is the transmission delay from associated satellite $s$ to the hosting satellite $s'$ for the first NF, while the other terms reflects the latency between satellites for the rest NFs in the required service. Similarly, the propagation delay can be presented as
\begin{equation}
	\label{eq:delay_PISL}
		\begin{aligned}
		T^{P(\mathbf{S}\rightarrow\mathbf{S'})}_u\!=\!\sum_{j\in\mathbf{J}}\!\sum_{k,k'\in\mathbf{K}}\!\sum_{s,s'\in\mathbf{S}}\mathbb{R}_{uj}\Bigl(\mathbb{A}_{us}\mathbb{V}_{1jk}\cdot y_{uks'}\!+\!\mathbb{A}_{us}\mathbb{V}_{Ijk}\!\cdot y_{uks'}\!+\!\sum_{i=1}^{I-1}(\mathbb{V}_{ijk}y_{uks})(\mathbb{V}_{(i+1)jk'}y_{uk's'})\Bigr)d_{s}h_{ss'}\frac{1}{c}
	\end{aligned}
\end{equation}
where the second term with subscript $i=I$ in eq.\eqref{eq:delay_PISL} illustrates the propagation delay of the return trip from the last NF hoster $s'$ to the associated satellite $s$, and $d_{ss'}$ expresses the distance between $s$ and $s'$.  
Furthermore, the computation time for terminal $u$'s request in satellites is expressed as 
\begin{equation}
	\label{eq:delay_CISL}
	T^{C(\mathbf{S})}_u=\sum_{j\in\mathbf{J}}\sum_{k\in\mathbf{K}}\sum_{s\in\mathbf{S}}\sum_{i=1}^{I}\frac{\mathbb{R}_{uj}\mathbb{V}_{ijk}c_kl_{uij}\cdot y_{uks}}{f_{ks}}
\end{equation}
where $f_{ks}$ is the allocated computation resources to NF $k$ in satellite $s$. 
In this paper, the edge computation resources are allocated according to the offloading decisions $y_{uks}$. When more than one requests are arranged to the same NF in a satellite, they will be served in a parallel manner,  which results in negligible queueing delay. 
In the end, the entire roundtrip latency is
\begin{equation}
	\label{eq:delay_RTISL}
	T^{RT(\mathbf{S}\rightarrow\mathbf{S'})}_u=T^{T(\mathbf{S}\rightarrow\mathbf{S'})}_u+T^{P(\mathbf{S}\rightarrow\mathbf{S'})}_u+T^{C(\mathbf{S})}_u
\end{equation}

\subsubsection{Satellite to Ground Station ($\mathbf{S}\rightarrow\mathbf{G}$)}The case of offloading computation task to ground station indicates that no satellite is selected to provide service for the terminal, i.e. $\sum_{s}y_{uks}=0$.  
By introducing $r_g$ as the satellite-to-terrestrial downlink capacity, we have the transmission delay as follows
\begin{equation}
	\label{eq:delay_Tsg}
	\begin{aligned}
	T^{T(\mathbf{S}\rightarrow\mathbf{G})}_u=&\sum_{j\in\mathbf{J}}\sum_{k\in\mathbf{K}}\frac{\mathbb{R}_{uj}\mathbb{V}_{1jk}l_{u1j}}{r_g}(1-\sum_{s\in\mathbf{S}}y_{uks})+\\
 &\sum_{j\in\mathbf{J}}\sum_{i=1}^{I-1}\frac{\mathbb{R}_{uj}l_{u(i+1)j}}{r_g}(1-\sum_{j'\in\mathbf{J}}\sum_{k\in\mathbf{K}}\sum_{s\in\mathbf{S}}\mathbb{R}_{uj'}\mathbb{V}_{ij'k}y_{uks})\oplus(1-\sum_{j'\in\mathbf{J}}\sum_{k\in\mathbf{K}}\sum_{s\in\mathbf{S}}\mathbb{R}_{uj'}\mathbb{V}_{(i+1)j'k}y_{uks})
	\end{aligned}
\end{equation} 
Similar to the delay expression eq.\eqref{eq:delay_TISL} of satellite-to-satellite case, the first term with subscript $i\!=\!1$ in eq.\eqref{eq:delay_Tsg} calculates the latency when a entire service chain is offloaded to the ground station, while the other terms represent the case of a partial service chain caching. We use $(1\!-\!\sum_{j,k,s}\mathbb{R}_{uj}\mathbb{V}_{ijk}y_{uks})$ to indicate whether the $i^\text{th}$ NF for terminal $u$ is served in the data center or not, and $\oplus$ as the exclusive OR (XOR) operator. Therefore, $(1\!-\!\sum_{j,k,s}\mathbb{R}_{uj}\mathbb{V}_{ijk}y_{uks})$$\oplus$$(1\!-\!\sum_{j,k,s}\mathbb{R}_{uj}\mathbb{V}_{(i+1)jk}y_{uks})$ becomes $1$ iff the first $i$ NFs are allocated to LEO satellites and the rest functions are redirected to the ground station. 
With the assumption that the resources in the data center is sufficient to serve all terminals simultaneously~\cite{tang2021iotj_computation}, we omit the queueing delay in the ground station. Regarding the computation latency, we have 
\begin{equation}
	\label{eq:delay_Csg}
		T^{C(\mathbf{G})}_u=\sum_{j\in\mathbf{J}}\sum_{k\in\mathbf{K}}\sum_{i=1}^{I}\frac{\mathbb{R}_{uj}\mathbb{V}_{ijk}c_kl_{uij}}{f_g}(1-\sum_{s\in\mathbf{S}}y_{uks})
\end{equation}
For the propagation delay, there is
\begin{equation}
	\label{eq:delay_Psg}
	T^{P(\mathbf{S}\rightarrow\mathbf{G})}_u\!=\!\frac{2d_g}{c}\cdot \mathbb{I}\left(\sum_{j\in\mathbf{J}}\sum_{k\in\mathbf{K}}\sum_{i=1}^{I}\mathbb{R}_{uj}\mathbb{V}_{ijk}(1\!-\!\sum_{s\in\mathbf{S}}y_{uks})\right)
\end{equation}
$\mathbb{I}(\cdot)$ is an indicator function that returns $1$ when the independent variable is true.  
Therefore, the entire roundtrip delay of the satellite-to-terrestrial link can be expressed as
\begin{equation}
	\label{eq:delay_RTsg}
	T^{RT(\mathbf{S}\rightarrow \mathbf{G})}_u=	T^{T(\mathbf{S}\rightarrow \mathbf{G})}_u+T^{C(\mathbf{G})}_u+T^{P(\mathbf{S}\rightarrow\mathbf{G})}_u
\end{equation}

Finally, the latency cost model for terminal $u$ becomes
\begin{equation}
	\label{eq:delay_total}
	T_u=T^{RT(\mathbf{U}\rightarrow \mathbf{S})}_u+	T^{RT(\mathbf{S}\rightarrow \mathbf{S'})}_u+	T^{RT(\mathbf{S}\rightarrow \mathbf{G})}_u
\end{equation}

\subsection{Energy Consumption Model}
 
With the assumption that the energy supply of the ground station is adequate\cite{zhang2023satellite}, we ignore the energy expenditure in the data center computation but focus on the consumption in terminals and satellites, which is explained as follows. 

\subsubsection{Terminal to Satellite ($\mathbf{U}\rightarrow \mathbf{S}$)} 
Let $p_{u}$ be the transmission power of the uplink between terminal and satellite. Then, the energy consumption of the terrestrial-to-satellite transmission is 
\begin{equation}
	\label{eq:energy_uplink}
	E^{T(\mathbf{U}\rightarrow \mathbf{S})}_u=p_{u}	T^{T(\mathbf{U}\rightarrow \mathbf{S})}_u
\end{equation}

\subsubsection{Satellite to Satellite ($\mathbf{S}\rightarrow\mathbf{S'}$)}We employ $p_s$ as the one-hop power consumption between satellites via ISL
and the related energy cost is
\begin{equation}
	\label{eq:energy_ISL}
	E^{T(\mathbf{S}\rightarrow \mathbf{S'})}_u=p_s T^{T(\mathbf{S}\rightarrow \mathbf{S'})}_u
\end{equation}
To illustrate the power consumption of task computation, we use the widely adopted model~\cite{zhang2023satellite,yang2022cost,song2021energy}, which is derived from the advanced dynamic voltage and frequency scaling technique, as follows
\begin{equation}
 	P^C_{ks}=\kappa f^3_{ks}
\end{equation}  
where $\kappa$ is a chip architecture-dependent coefficient. With the computation time in eq.\eqref{eq:delay_CISL}, the energy consumption for task computation can be calculated by 
\begin{equation}
	E^{C(\mathbf{S})}_u=\kappa\sum_{j\in\mathbf{J}}\sum_{k\in\mathbf{K}}\sum_{s\in\mathbf{S}}\sum_{i=1}^{I}\mathbb{R}_{uj}\mathbb{V}_{ijk}c_kl_{uij}\cdot f^2_{ks}y_{uks}
\end{equation}
In this part we do not take the energy for hosting NFs into consideration. The rational behind this is that edge platforms on LEO satellite are assumed to be kept in power-on mode to provide various kinds of services, such as routing and on-board computing, instead of only NF caching.
\subsubsection{Satellite to Ground Station ($\mathbf{S}\rightarrow\mathbf{G}$)}
By using $p_g$ as the satellite-to-terrestrial downlink transmission power, we have  
\begin{equation}
	\label{eq:energy_downlink}
	E^{T(\mathbf{S}\rightarrow\mathbf{G})}_u=p_g T^{T(\mathbf{S}\rightarrow\mathbf{G})}_u
\end{equation}

Therefore, the entire energy consumption model for terminal $u$ can be denoted as
\begin{equation}
	\label{eq:energy_total}
	E_u=E^{T(\mathbf{U}\rightarrow\mathbf{S})}_u+	E^{T(\mathbf{S}\rightarrow\mathbf{S'})}_u+	E^{T(\mathbf{S}\rightarrow\mathbf{G})}_u+E^{C(\mathbf{S})}_u
\end{equation}

\section{Problem Formulation}
\label{sec:MILP}
\subsection{Optimization Model}
For each LEO satellite, the allocated computation resource should not exceed the edge computing capability, which is expressed as
\begin{equation}
	C_1:~\sum_{u\in\mathbf{U}}\sum_{k\in\mathbf{K}}f_{ks}y_{uks}\leq C_s, \forall s\in\mathbf{S}
\end{equation}

Similarly, the utilized caching space should be limited within the storage capacity, which is denoted as
\begin{equation}
	C_2:~\sum_{k\in\mathbf{K}}l_{k}x_{ks}\leq L_s, \forall s\in\mathbf{S}
\end{equation}

Additionally, the selected satellite for computation offloading should cache the relevant NF, i.e. 
\begin{equation}
	C_3:~y_{uks}\leq x_{ks}, \forall u\in\mathbf{U},k\in\mathbf{K}, s\in\mathbf{S}
\end{equation} 

Furthermore, in each time slot, the terminal $u$'s request of NF $k$ cannot be served by more than one satellite
\begin{equation}
	C_4:~\sum_{s\in\mathbf{S}}y_{uks}\leq 1, \forall u\in\mathbf{U},k\in\mathbf{K}
\end{equation}

As aforementioned, when the request of the $i^\textit{th}$ NF $k$ of terminal $u$ is offloaded to the date center, i.e. $\sum_sy_{uks}=0$, the following following NF requests ($i+1$,$\cdots$,$I$) would be served in the same date center instead of in LEO satellites. In order to simplify notations, we define 
\begin{equation}
	g_{iu}(y)=\sum_{j\in\mathbf{J}}\!\sum_{k\in\mathbf{K}}\!\sum_{s\in\mathbf{S}}\!\mathbb{R}_{uj}\!\mathbb{V}_{ijk}y_{uks} 
\end{equation}
Then the limitation can be expressed by the following constraint
\begin{equation}
C_5:g_{iu}(y)\geq M\cdot g_{(i+1)u}(y), \forall u\in\mathbf{U},i=1,\cdots,I-1
\end{equation}
where $M$ is a sufficiently large number. 

The total cost is a combination of delay and energy for all terminals, which is
\begin{equation}
	\label{eq:total_cost}
	\Delta=\sum_{u\in\mathbf{U}}\left(\alpha\cdot \tilde{T}_u+(1-\alpha)\cdot\tilde{E}_u\right)
\end{equation} 
where $\tilde{T}_u$ and $\tilde{E}_u$ are the latency cost $T_u$ in eq.\eqref{eq:delay_total} and entire energy consumption $E_u$ in eq.\eqref{eq:energy_total} after normalization, respectively, for the purpose of multi-objective optimization; $\alpha$ is the weight for the delay-energy tradeoff with $\alpha\in[0,1]$.

Then, the cost-efficient service chain caching and computation offloading problem can be jointly formulated as
\begin{equation}
	\label{fml:op1}
	\begin{aligned}
		(\mathbf{P1}):\quad\mathop{\min}_{\mathbf{X},\mathbf{Y}}\;&\Delta\\
		\textrm{s.t.}\quad&C_1-C_5\\
		&\mathbf{X},\mathbf{Y}\in\{0,1\}
	\end{aligned}
\end{equation}

\subsection{Model Linearization}	
Note that both the expression of $T^{T(\mathbf{S}\rightarrow\mathbf{S'})}_u$ in eq.\eqref{eq:delay_TISL} and $T^{P(\mathbf{S}\rightarrow\mathbf{S'})}_u$ in eq.\eqref{eq:delay_PISL} contain a product of two binary decision variables, i.e. $y_{uks}\cdot y_{uk's'}$, thus we introduce an auxiliary decision variable $\mathbf{Z}=\{z_{uksk's'}|z_{uksk's'}\!=\!\{0,1\},u\!\in\!\mathbf{U},k,\!k'\in\mathbf{K},s,\!s'\in\mathbf{S}\}$, which is defined as 
\begin{equation*}
	z_{uksk's'}=y_{uks}\cdot y_{uk's'}=
	\begin{cases}
		1,&\textit{if}~y_{uks}=1~\textit{and}~y_{uk's'}=1;\\
		0,&\textit{otherwise}.
	\end{cases}
\end{equation*}
The following constraints are equivalent to the definition of $z_{uksk's'}$ by examining the combinations of $y_{uks}$ and $y_{uk's'}$. 
\begin{align}
	C_6:~&z_{uksk's'}\!\leq\!y_{uks}, \forall u\!\in\!\mathbf{U},k,\!k'\!\in\!\mathbf{K},s,\!s'\!\in\!\mathbf{S}\\
	C_7:~&z_{uksk's'}\!\leq\!y_{uk's'}, \forall u\!\in\!\mathbf{U},k,\!k'\!\in\!\mathbf{K},s,\!s'\!\in\!\mathbf{S}\\
	C_8:~&z_{uksk's'}\!\geq\!y_{uks}\!+\!y_{uk's'}\!-\!1, \forall u\!\in\!\mathbf{U},k,\!k'\!\in\!\mathbf{K},s,\!s'\!\in\!\mathbf{S}
\end{align}

In eq.\eqref{eq:delay_Tsg},  $T^{T(\mathbf{S}\rightarrow\mathbf{G})}_u$ includes an XOR operation. An auxiliary variable $\mathbf{\Theta}=\{\theta_{iu}|\theta_{iu}\in\{0,1\},u\!\in\!\mathbf{U},i\!=\!1,\cdots,I-1\}$ is utilized with definition
\begin{equation}
\theta_{iu}=(1-g_{iu}(y))\oplus(1-g_{(i+1)u}(y))
\end{equation}
And then $\mathbf{\Theta}$ must satisfy the following constraints
\begin{align}
&C_9\!:\theta_{iu}\!\geq\! g_{iu}(y)\!-\!g_{(i+1)u}(y), \forall u\!\in\!\mathbf{U},i\!=\![1,I\!-\!1]\\
&C_{10}\!:\theta_{iu}\!\geq\! g_{(i+1)u}(y)\!-\!g_{iu}(y), \forall u\!\in\!\mathbf{U},i\!=\![1,I\!-\!1]\\
&C_{11}\!:\theta_{iu}\!\leq\! 2\!-\!g_{iu}(y)\!-\!g_{(i+1)u}(y), \forall u\!\in\!\mathbf{U},\!i\!=\![1,I\!-\!1]\\
&C_{12}\!:\theta_{iu}\!\leq\! g_{iu}(y)\!+\!g_{(i+1)u}(y), \forall u\!\in\!\mathbf{U},i\!=\![1,I\!-\!1]
\end{align}

Moreover, to eliminate the indicator expression in eq.\eqref{eq:delay_Psg}, we employ another auxiliary decision variable $\mathbf{\Pi}=\{\pi_u|\pi_u=\{0,1\},u\in\mathbf{U}\}$ as the indicator of offloading to the ground station, which can be determined by 
\begin{equation}
	\label{eq:indicator}
	\pi_u=\mathbb{I}\left(\sum_{j\in\mathbf{J}}\sum_{k\in\mathbf{K}}\sum_{i=1}^{I}\mathbb{R}_{uj}\mathbb{V}_{ijk}(1\!-\!\sum_{s\in\mathbf{S}}y_{uks})\right)
\end{equation}
Similar to \cite{wang2020sfcbased}, we can transform eq.\eqref{eq:indicator} into 
\begin{equation}
	C_{13}\!:\!\sum_{j\in\mathbf{J}}\!\sum_{k\in\mathbf{K}}\!\sum_{i=1}^{I}\mathbb{R}_{uj}\!\mathbb{V}_{ijk}\!(1\!-\!\sum_{s\in\mathbf{S}}y_{uks})\!<\!M\!\pi_u\!+\!1, \forall u\!\in\!\mathbf{U}
\end{equation}

Finally, the optimization model ($\mathbf{P1}$) can be rewritten in the form of ILP model:
\begin{equation}
	\label{fml:op2}
	\begin{aligned}
		(\mathbf{P2}):\quad\mathop{\min}_{\mathbf{X},\mathbf{Y},\mathbf{Z},\mathbf{\Theta},\mathbf{\Pi}}\;&\Delta\\
		\textrm{s.t.}\quad&C_1-C_{13}\\
		&\mathbf{X},\mathbf{Y},\mathbf{Z},\mathbf{\Theta},\mathbf{\Pi}\in\{0,1\}
	\end{aligned}
\end{equation}
As for the time complexity, the proposed model falls into the family of NP-hard. The proof can be established by setting each service chain contains only one NF. Then, by viewing the LEO satellite, service, and terminal requests as locations, facilities, and demands respectively, ($\mathbf{P2}$) can be converted to the Facility Location Problem (FLP). Therefore ($\mathbf{P2}$) is at least as difficult as the well-known NP-hard problem FLP, and consequently ($\mathbf{P2}$) also belongs to NP-hard.

\subsection{Heuristic Algorithm}
Due to the curse of dimensionality, solving ($\mathbf{P2}$) is very time-consuming, especially when the network instance becomes larger. To tackle this dilemma, we introduce the Greedy Caching and Offloading (GCO) algorithm which attempts to allocate each NF request to its nearest LEO satellite, as detailed in Alg.\ref{alg:heuristic}. Similar to the content caching decisions, we evaluate the popularity of a NF by calculating the number of being requested, i.e. 
\begin{equation}
	\label{eq:popul}
	N_{iks}=\sum_{j\in\mathbf{J}}\sum_{u\in\mathbf{U}}\mathbb{A}_{us}\mathbb{R}_{uj}\mathbb{V}_{ijk}
\end{equation}  
It is worth noting that in eq.\eqref{eq:popul} the order $i$ is also taken into account. We start caching and computation offloading for the NF which is the first in a service chain and has the most requests. There are five loops in Alg.\ref{alg:heuristic}, and, thus, the time complexity of GCO is $O(I\vert\mathbf{K}\vert\vert\mathbf{U}\vert\vert\mathbf{S}\vert^2)$. Once the network topology is determined in the considered time epoch, $\vert\mathbf{S}\vert$ becomes constant and, as a result, the running time is reduced to $O(I\vert\mathbf{K}\vert\vert\mathbf{U}\vert)$, i.e. cubic time complexity.

\begin{algorithm}[!htbp]
	\caption{Greedy Caching and Offloading Algorithm~(GCO)}
	\label{alg:heuristic}
	\KwData{Variables in Table \ref{tab:notations}}
	\KwResult{Cache allocation $\mathbf{X}$ and offloading decision $\mathbf{Y}$}
	Initialize $\mathbf{X}\leftarrow\mathbf{0}$ and $\mathbf{Y}\leftarrow\mathbf{0}$\;
	Calculate the popularity of NF $k$ in LEO $s$ via eq.\eqref{eq:popul}\;
	\ForEach{$i=1,\cdots,I$,$s\in\mathbf{S}$}{
		Set $\mathbf{K'}$ $\leftarrow$ sort $k$ based on $N_{iks}$ descendingly\;
		\ForEach{$k\in\mathbf{K'}$,$u\in\mathbf{U}$}{
			\If{NF $k$ is requested by $u$}{
				\If{$L_s\leq l_k$ \textbf{and} $x_{ks}=0$}{
				$x_{ks}\leftarrow 1$, $L_s\rightarrow L_s-l_k$\;
				}
				Set $\mathbf{S'}$ $\leftarrow$ sort $s'$ based on $h_{ss'}$ ascendingly\;
				\ForEach{$s'\in\mathbf{S'}$}{
				\If{$C_{s'}\leq f_{ks'}$ \textbf{and} $x_{ks'}=1$ \textbf{and} $y_{uks'}=0$}{
						$y_{uks'}\leftarrow 1$, $C_{s'}\rightarrow C_{s'}-f_{ks'}$\;
						\textbf{break}\;
						}					
				}
			}
		}
	}
\end{algorithm}

\section{Numerical Investigations}

\subsection{Simulation Settings}
In this section, we demonstrate the performance of the optimal decision making derived from the proposed ILP model ($\mathbf{P2}$,), which is named as ILP in the following,  together with the defined GCO algorithm. To further benchmark the results we also compare with Data center Caching and Offloading (DCO) method and Network Function Caching and Offloading (NFCO) algorithm. As the name indicates, the LEO satellites in DCO act as relays and all NF requests would be redirected to the data center. When calculating the total cost $\Delta$ in eq.\eqref{eq:total_cost}, the delay and energy consumption of DCO are employed as the normalization factor for $\tilde{T}_u$ and $\tilde{E}_u$, respectively. Based on the available storage space and computing resource, NFCO allocates NF caching and offloading to the nearby LEO satellites. In the NFCO, caching decisions for various NFs are decided independently and the operation order of NFs in service chains is not considered. 

The experiments are performed in a 8-LEO satellites network model, and each satellite establishes ISLs with four adjacent satellites. All results presented hereafter are averaged over 100 various simulation scenarios. 
The simulation parameters are summarized in Tab.\ref{tab:sim}. Unless otherwise provided, we would set $\vert\mathbf{U}\vert$, $r_u$, $C_s$, and $f_{ks}$ as their default values.

\begin{table}[htb]
	\centering
	\caption{Simulation Parameters\cite{zhang2023satellite,yang2022cost,wang2018computation}}
	\label{tab:sim}
	\begin{tabular}{lcc}
		\hline
		\textbf{Parameter}&\textbf{Default Value}&\textbf{Range}\\
		\hline
		Number of NFs in chain $I$ & 4 &-\\
		Transmit power $p_u$, $p_s$, $p_g$ &\{3, 30, 20\}dBW&-\\
		Distance $d_u$,$d_s$, $d_g$&\{1000,800,2000\}Km&-\\
		Date rate $r_s$, $r_g$ & \{10,0.3\}Gbps&-\\
		Input data size $l_{uij}$ & - & [1,100]Mb\\
		NF work load $c_k$ & 100 cycles/bit & -\\
		Allocated resources $f_g$&2 Gcycles/s &-\\
		Number of terminals $\vert\mathbf{U}\vert$ & 10 & [5,30]\\
		Date rate $r_u$ & 200Mbps & [50,350]Mbps\\
		Computation capacity $C_s$&10 Gcycles/s&[2,22]Gcycles/s\\
		Allocated resources $f_{ks}$&2 Gcycles/s& [0.5,3]Gcycles/s\\
		\hline
	\end{tabular}
\end{table}

\subsection{Performance Comparison}

\begin{figure*}[htpb]
	\centering
	\begin{subfigure}{.48\textwidth}
		\centering
		\includegraphics[width=\linewidth]{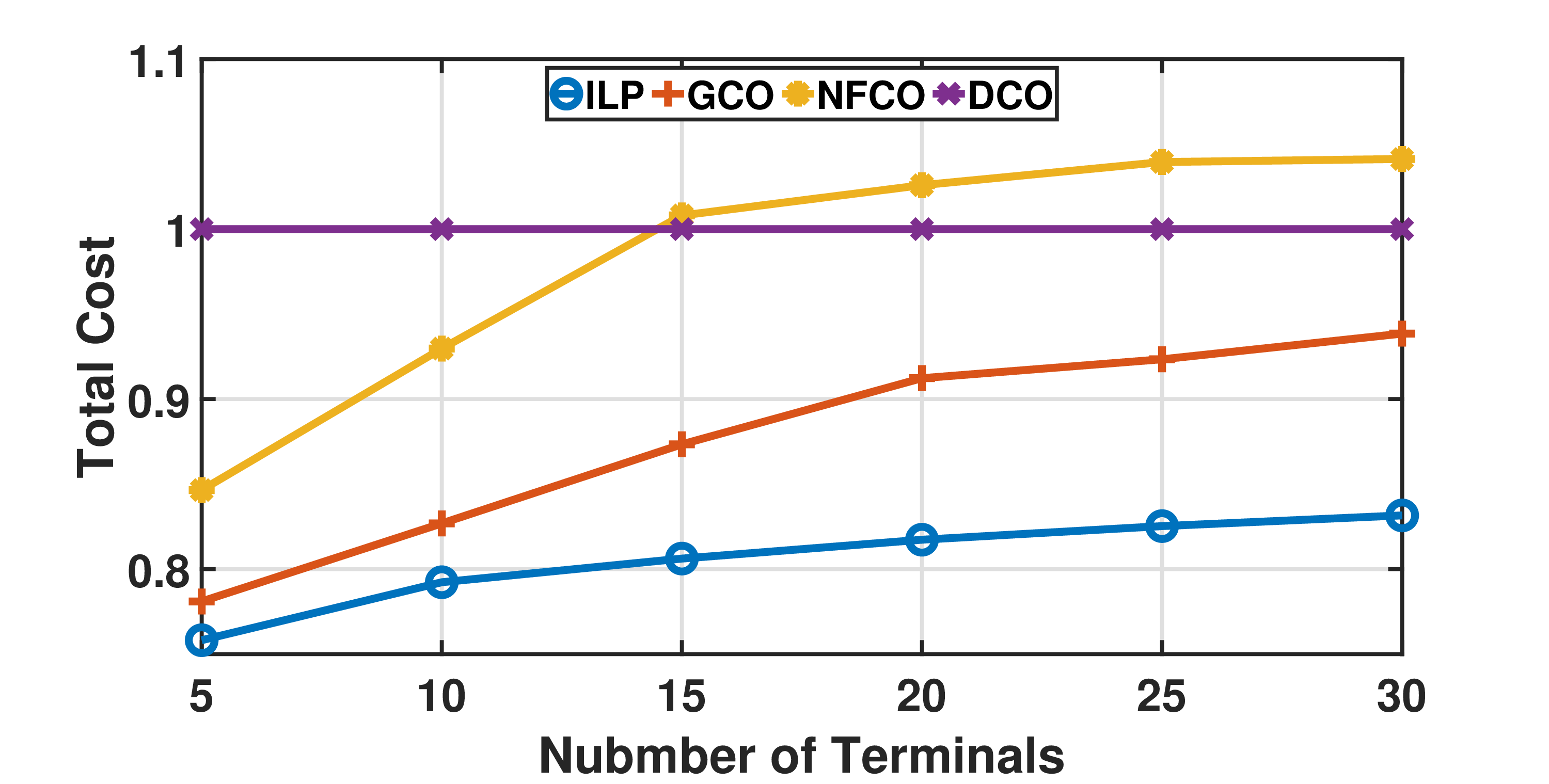}
		\caption{Total cost}
		\label{fig:tc_u}
	\end{subfigure}
	\begin{subfigure}{.48\textwidth}
		\centering
		\includegraphics[width=\linewidth]{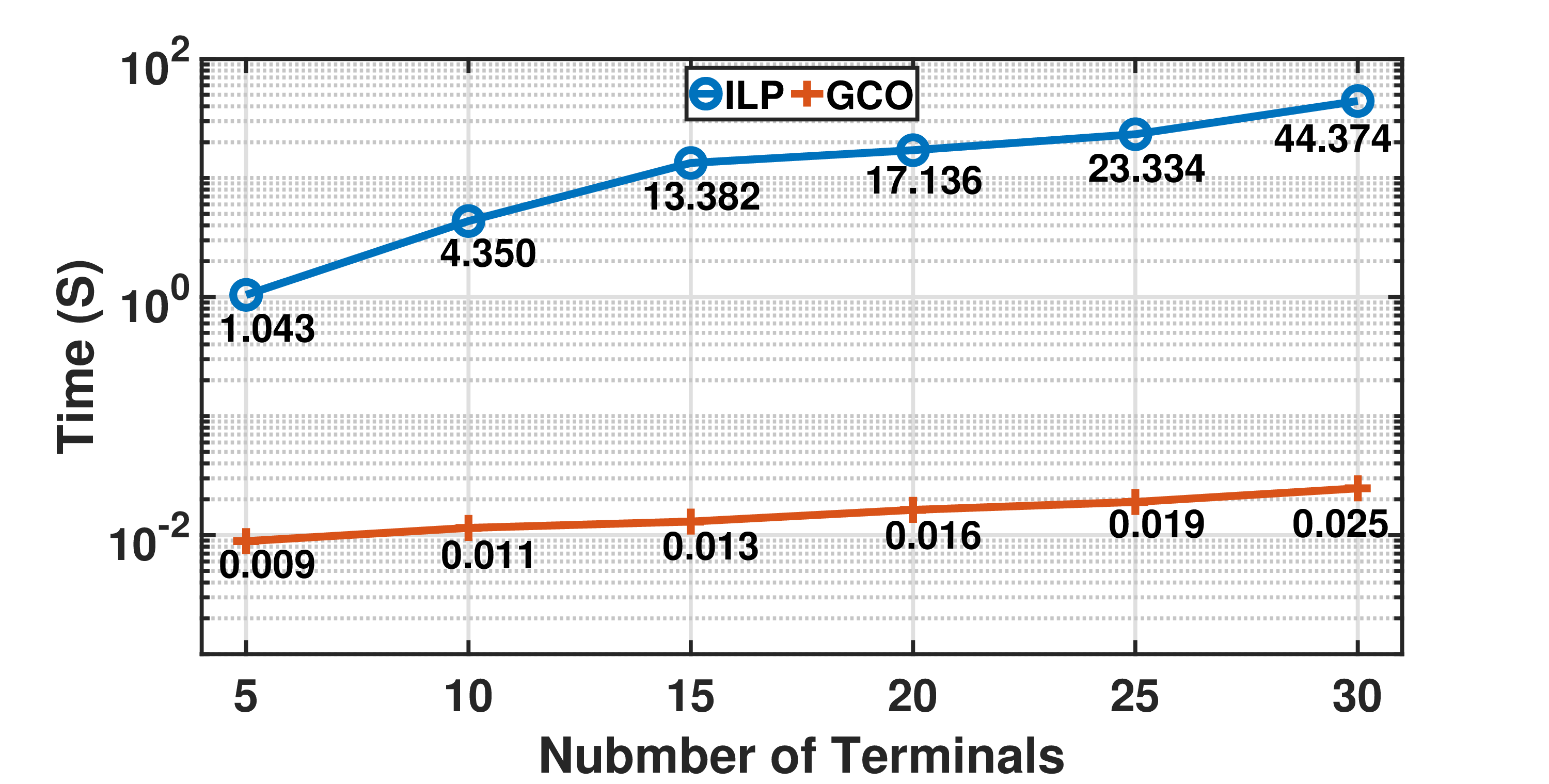}
		\caption{Time complexity}
		\label{fig:time_u}
	\end{subfigure}
	\caption{\label{fig:U}Performance comparison with various number of terminals}
\end{figure*}

We first study how the number of terminals affects the performance of proposed algorithms from aspects of normalized total cost and time complexity respectively. As aforementioned, we utilize the result of DCO as the normalized factor to calculate total cost. Therefore, the DCO curve remains steady in Fig.\ref{fig:tc_u}. However, the results of ILP, GCO, and NFCO are sensitive to the change of terminals. The increase in terminals not only introduces additional noise in communication channels, which results in less uplink data rate, but also makes the competition of CPU fiercer. As a result, the resources in LEO satellite system  become saturated soon, and, then, offloading NFs to the data center results in additional expenditure, which reflects 3 rising curves in Fig.\ref{fig:tc_u}. When the number of terminals ranges between $5$ and $15$, even not taking the specific operation order into consideration, we can also benefit from caching and offloading to LEO. The gain ranges from 24.2\% to 19.9\% for the ILP, 21.9\% to 13.6\% for the GCO, and 15.2\% to 0\% for the NFCO. After this turning point, due to the independent decision making among NFs, the extra cost paid for data transmissions among satellites and ground stations exceeds the cost saving from caching and offloading to LEO. Consequently, the NFCO consumes more cost than the DCO. 

The running time of the ILP and GCO are compared in Fig.\ref{fig:time_u}. For the former, the time spent on solving ILP model varies from around 1 second to more than half-minute, while latter can provide results within 30 ms. 
Albeit with higher computational complexity, the ILP outperforms the other 3 methods. The heuristic algorithm GCO is able to make competitive decisions, especially when the number of terminals is less than 15, and, then, NFCO and DCO follow. For instance, compared to the ILP, GCO saves more than $99$\% running time by extra 5\% total cost payment. 

\begin{figure*}[htpb]
	\centering
	\begin{subfigure}{.48\textwidth}
		\centering
		\includegraphics[width=\linewidth]{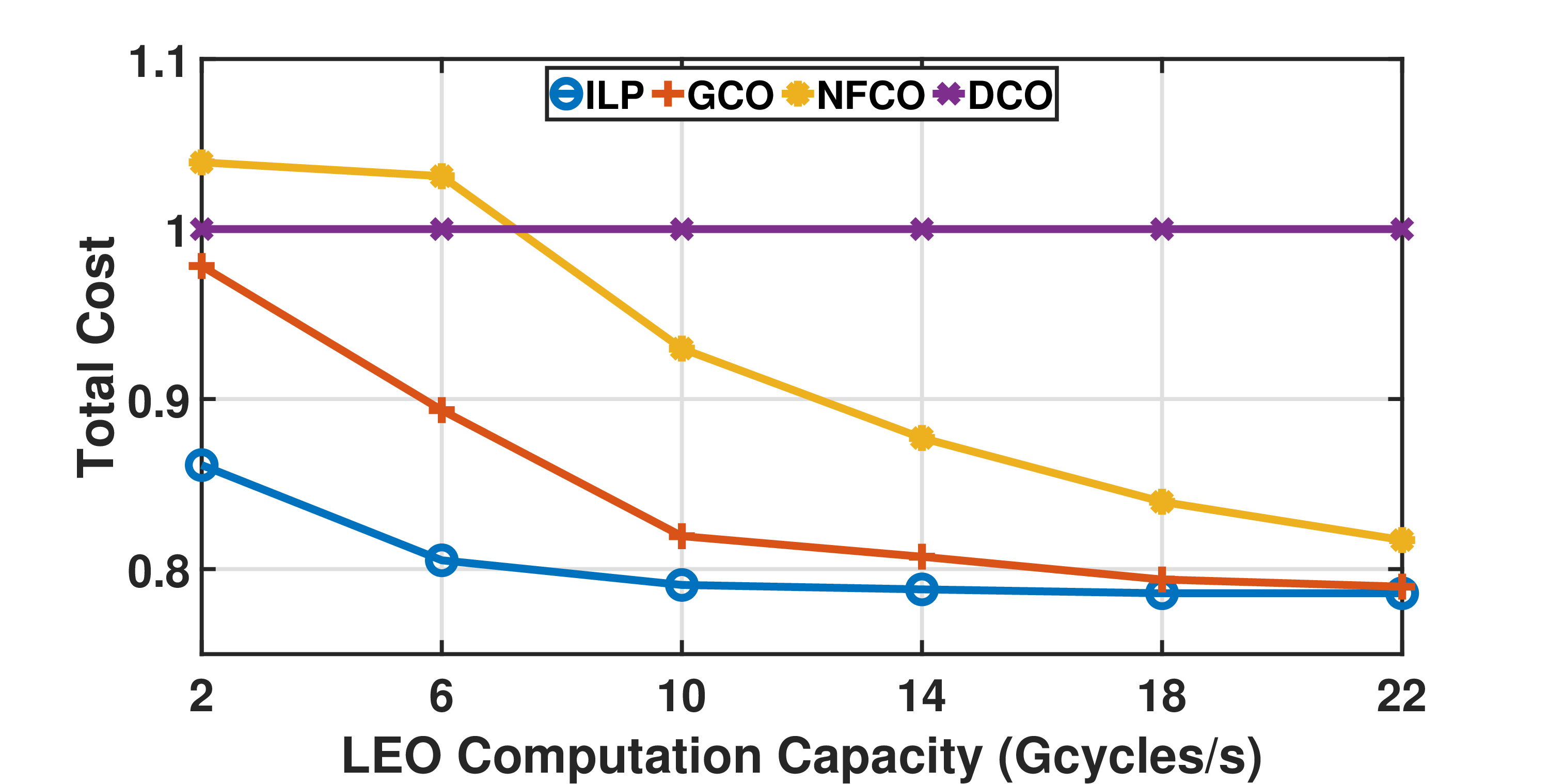}
		\caption{The effect of LEO computing resources}
		\label{fig:LEO_com}
	\end{subfigure}
	\begin{subfigure}{.48\textwidth}
		\centering
		\includegraphics[width=\linewidth]{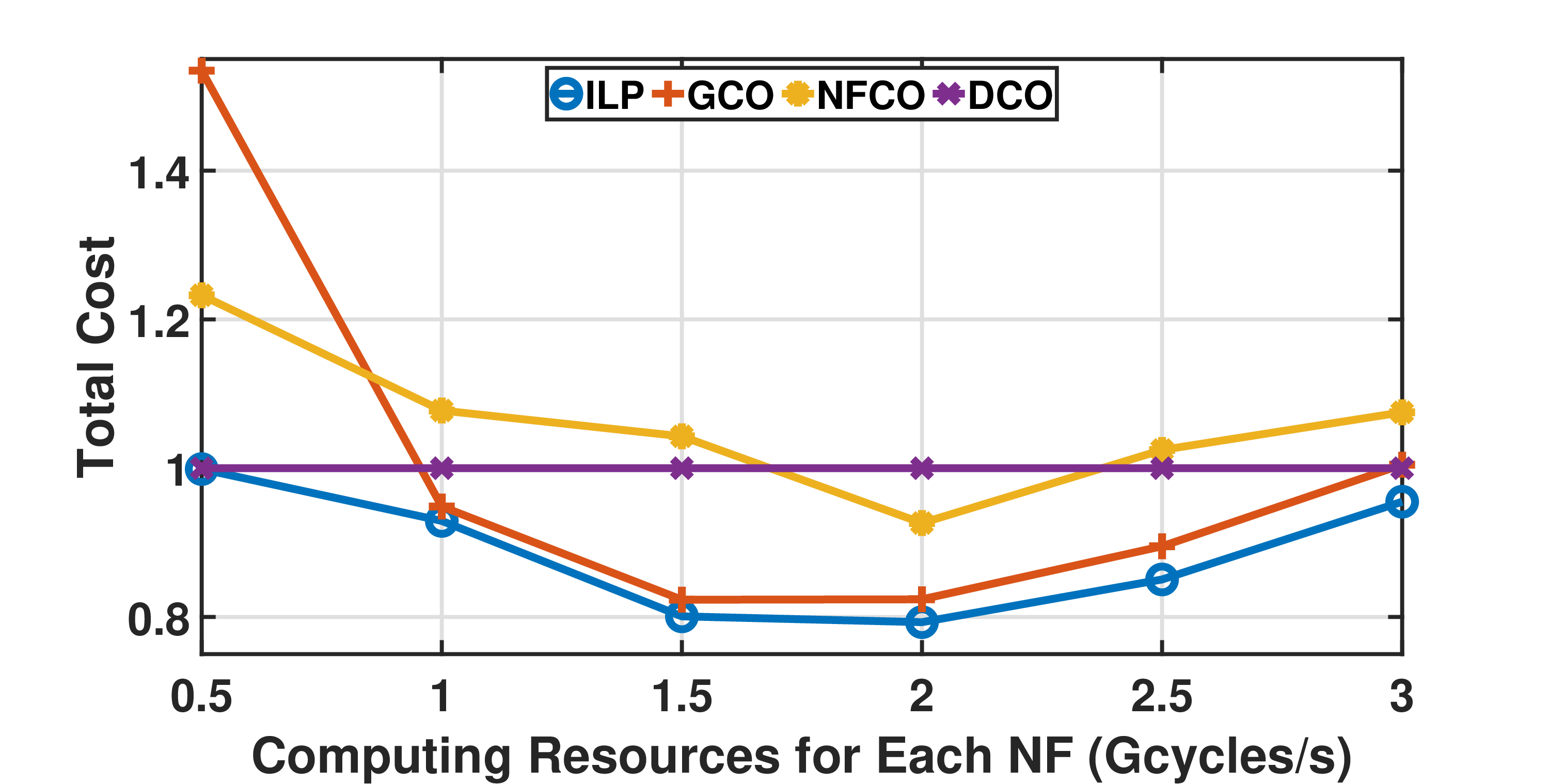}
		\caption{The effect of resource allocation}
		\label{fig:LEO_allo}
	\end{subfigure}
	\caption{\label{fig:LEO}Performance comparison with various capabilities and allocations}
\end{figure*}

Fig.\ref{fig:LEO} investigates the influence of various LEO capability settings. As shown in Fig.\ref{fig:LEO_com}, the benefit from caching and on-board computation offloading in LEO satellites is very limited without the support of sufficient LEO computation capacity. By increasing the available CPU resources from 2 to 10 Gcycles/s, there is a rapid decrease in the total cost of GCO and ILP. However, expanding the computation capacity continuously does not provide a gradual improvement on the performance. This result can be explained by considering the constraint of storage spaces on LEO, which is limited by ($C_3$). 
The effect of computation resources allocation to each NF in satellite is tested in Fig.\ref{fig:LEO_allo}. On the one hand, an appropriate increase in the CPU resource allocation to a NF accelerates the computation process in LEO, which results in less latency and computing energy consumption; On the other hand, an excessive allocation to a NF exhausts the computation resources rapidly and, thus, reduces the number of terminals being served in LEO satellites. As expected, Fig.\ref{fig:LEO_allo} reveals that the curves of ILP, GCO, and NFCO share a dropping-climbing trend with the growth of $f_{ks}$. This result indicates that more attention should be paid to managing the computation resources allocation in order to fully capitalize the benefits from LEO caching and offloading.

\section{Conclusions}

Providing effective caching and offloading strategies in low earth orbit satellites that incorporate latency and energy consumption considerations is significant to support emerging applications that require ubiquitous and continuous services. In this article, we jointly consider the problem of anchoring network functions and deciding proper satellites for computation services, which is formulated and linearized as an integer linear programming model. Specifically, the operation order of functions in service chains is also taken into account. In addition a greedy algorithm is presented and the effectiveness of proposed methods is evaluated with a wide set of numerical investigations. 
Some future directions are envisioned as follows. In this paper, we focus on one time epoch with a quasi-static model. Further analysis and discussion are needed in the case of multi time slots including dynamic network models. Furthermore, one can study the joint optimization of caching, computing, and communicating resources allocation in non-terrestrial system, so as to capitalize the resource utilization. Finally, artificial intelligence-driven algorithms can be considered to augment the model with learning capabilities, which improves the adaptation to various network conditions.

\bibliographystyle{ieeetr}
\bibliography{references} 
\end{document}